# Achieving digital-driven patient agility in the era of big data

Rogier van de Wetering

[1] Faculty of Sciences, Open University of the Netherlands, Valkenburgerweg 177, 6419 AT Heerlen, the Netherlands rogier.vandewetering@ou.nl

**Abstract.** There is still a limited understanding of the necessary skill, talent, and expertise to manage digital technologies as a crucial enabler of the hospital's ability to adequately 'sense' and 'respond' to patient needs and wishes, i.e., patient agility. Therefore, this investigates how hospital departments can leverage a 'digital dynamic capability' to enable the department's patient agility. This study embraces the dynamic capabilities theory, develops a research model, and tests it accordingly using data from 90 clinical hospital departments from the Netherlands through an online survey. The model's hypothesized relationships are tested using structural equation modeling (SEM). The outcomes demonstrate the significance of digital dynamic capability in developing patient sensing and responding capabilities that, in turn, positively influence patient service performance. Outcomes are very relevant for the hospital practice now, as hospitals worldwide need to transform healthcare delivery processes using digital technologies and increase clinical productivity.

## 1 Introduction

In the age of big data and data-driven decision-making, hospitals worldwide use innovative information technology (IT) to transform their care delivery processes and business models, thereby improving cost efficiency, clinical quality, service efficiency, and patient satisfaction [1-7]. Hospitals are forced to do so, as they are an essential component of modern-day society, and the adoption of groundbreaking IT essential to its success [7-9]. Hospitals use IT, such as clinical decision-support systems (CDSS), to enhance the decision-making processes and provide clinicians with several modes of decision support (e.g., alerts, reminders, advice) [10-12]. Another recent development is using big data and predictive analytics as doctors need to analyze exponential volumes of patient-generated data [13]. Big data, in essence, refers to datasets whose size is beyond the ability of conventional database software tools to capture, store, manage, and analyze both structured and unstructured data. Modern hospitals are currently very active in exploring new digital options and data-driven innovations using big data to drive clinical care quality and strengthen the relationships and interactions with patients. For instance, clinicians use digital innovations in their clinical practice, e.g., mobile handheld devices and apps, to increase error prevention and improve patient-centered care [14]. Also, digital innovations provide clinicians with ways to be more agile in their work, improve clinical communication, remotely monitor patients, enhance clinical decision-support [8, 9], and improve the patient treatment process and medical quality services [15]. Big data analytics is particularly relevant for hospitals where it can be used to, e.g., to identify defects in care and risk factors for patient safety issues, determine the time required to perform key patient care activities (e.g., passing medication), occurrence patterns, and statistical testing of intervention strategies, evidence-based medicine, and analyses to measure the impact of using clinically substitutable supply items on patient outcomes.

There is a wealth of attention for information technology (IT) adoption and IT-enabled transformation in healthcare research [16]. However, there is still a limited understanding of big data's role and its associated predictive models as a crucial enabler of the hospital's ability to adequately 'sense' and 'respond' to patient needs and wishes, i.e., patient agility [13, 17] and hospitals have not fully grasped the value of these data-driven innovations yet [13]. As such, this article embraces the dynamic capabilities theory (DCT), a foundational strategic framework within the management and IS field, when it comes to the innovative and orchestrated use of digital technologies [18-21]. When hospital departments want to excel and use data-driven innovations in practice and drive patient agility, for instance, to help detect COVID-19 cases early using big data, they need to manage and master digital technologies. Hence, they need to develop a 'digital dynamic capability' which can be considered the "organization's skill, talent, and expertise to manage digital technologies for new innovative product development" [22]. For such a capability to develop, the hospital department needs heterogeneous competencies [17, 23]. Against this background and the current gaps in the literature, this paper acclaims that digital dynamic capability enhances the ability to sense and respond adequately to the patient's needs and demands and drive the department's patient service performance, i.e., the extent to which hospital departments achieve high-quality medical services [24]. Hence, this research addresses the following research questions:

*(I) What is the effect of digital dynamic capability on the hospital departments' sense-and-respond capabilities, i.e., patient agility? Also, (II) through what mechanism does patient agility lead to high levels of patient service performance?*

## 2 Theoretical foundation

This study builds upon the dynamic capabilities theory (DCT) [23, 25]. The DCT is a leading theoretical framework that explains where firms' competitive advantage comes from in industries with high technological and market turbulence. Dynamic capabilities can be defined as a specific subset of capabilities that allow firms to integrate, build, and reconfigure internal and external resources and competences to create new products and processes and respond to changing business environments [26]. Hence, these capabilities allow firms to manage uncertainty [19, 25]. Notwithstanding its significance, the theory has been profoundly subjected to theoretical debate [23, 25-27]. However, most empirical endeavors established positive relationships among these capabilities in recent years, firm's operational, innovative and competitive performance measures [28].

The concept of organizational agility is a manifested type of dynamic capability [19]. It can be conceptualized as a dynamic capability if "*they permit organizations to repurpose or reposition their resources as conditions shift"* [29]. Organizational agility has been proposed under the DCT as an essential organizational capability to respond to changing conditions while simultaneously proactively enacting the dynamic environment regarding customer demands, supply chains, new technologies, governmental regulations, and competition [19, 30, 31]. This 'sense-and-respond' capability has been defined and conceptualized in many ways and through various theoretical lenses in the IS literature [32, 33]. For instance, Park et al. [31] ground their conceptualization and operationalization in the information-processing theory and argue that information processing capabilities strengthen the organization's sense-response processes to adapt to changing environmental conditions. Lu and Ramamurthy [34] embrace a complementarity perspective and perceive agility as the organization's ability to seize market opportunities and operationally adjustment capacity. Roberts and Grover [35] synthesized that, although there seems to be ambiguity in definitions as reflected by the concepts' operationalized capabilities, a set of high-level characteristics can be devised from the extant literature: deliberately 'sensing' and 'responding' to business events in the process of capturing business and market opportunities. This article perceived patient agility as a higher-order manifested type of dynamic capability that allows hospital departments to adequately 'sense' and 'respond' to patient-based opportunities, needs, demands within a fast-paced hospital ecosystem context [19, 35]. Digital dynamic capability is a crucial technical-oriented dynamic capability necessary to innovate and enhance business operations using digital technologies [22, 36-38]. These digital technologies include big data analytics and artificial intelligence (AI). Digital dynamic capability can be conceived as an organization's ability to master digital technologies, drive digital transformations, and adopt new innovative services and products. Digital dynamic capability is conceptualized as a lower-order technical dynamic capability that facilitates developing higher-order dynamic organizational capabilities such as innovation ambidexterity, absorptive capacity, and organizational agility [17, 27, 38, 39].

# 3 Research model and hypothesis development

## 3.1 Digital dynamic capability and patient agility

Recent scholarship shows that digital dynamic capability is crucial to innovate and enhance business operations [22, 36-38]. For instance, Wang et al. [40] argue that firms use the digital dynamic capability to leverage IT and knowledge resources to deliver innovative services that customers value. Coombs and Bierly [41] empirically showed that this technological-driven capability enables competitive advantages. The literature shows that by actively managing the opportunities provided by new digital innovation such as big data and AI and actively responding to digital transformation, organizations can succeed in their digital options and services [22, 40]. This capability is vital for hospital departments that want to strive for patient agility in clinical practice because the process of achieving new digital patient service solutions is exceedingly dependent on its ability to manage digital technologies [17, 22]. The digital dynamic capability provides the hospital department with the ability to, e.g., integrate devices (think, for instance, about patient location devices, smart beds, bed & tracking boards) so that accurate and efficient clinical documentation and processing is facilitated and better clinical decision takes places [6]. Hence, hospitals that actively invest and develop these capabilities are likely to sense and anticipate their patients' needs (of which they might be physically and mentally unaware) and respond fast to changes in the patient's health service needs using digital innovations and assessments of clinical outcomes [17, 22, 35]. Therefore, the following two hypotheses are defined:

**Hypothesis 1:** *Hospital departments' digital dynamic capability positively impacts the department's patient sensing capability.*

**Hypothesis 2:** *Hospital departments' digital dynamic capability positively impacts the department's responding capability.*

### 3.2 Patient sensing and responding capability

This article, thus, hypothesizes that hospital departments' digital dynamic capability is key to establishing patient sensing and responding capabilities. Furthermore, it is suggested that the digital-driven patient sensing capability, in turn, affects the hospital departments' ability to respond rapidly if something important happens with the patients or their service needs. Digital options and innovations provide clinicians with ways to sense and anticipate patient's needs, wishes, and demands more effectively [7, 42] and thereby improve the patient treatment process and quality of medical services [15, 42]. However, in order for a responding capability to be effective, the hospital department is dependent on its ability to sense and anticipate the patients' needs [35]. In a similar vein, it can be argued that hospital departments cannot leverage a patient responding capability if they have not developed an effective sensing capability. Hence, the following is defined:

**Hypothesis 3:** *Hospital departments' sensing capability positively impacts the department's patient responding capability.*

### 3.3 Patient responding capability and patient service performance

The extant literature shows that digital-driven sense and respond capabilities are crucial to achieving higher-quality and patient-centered care and hospitals' financial performance benefits [24, 43]. By making specific investments in capabilities valued by patients, hospital departments can achieve high levels of patient service performance and value in the turbulent healthcare environment [44]. For example, clinicians who use digital innovations in their clinical practice, e.g., mobile handheld devices and apps, can increase error prevention and improve patient-centered care [14, 43]. However, a sufficient responding capability should be preceded by a developed and aligned sensing capability to respond effectively and drive patient service performances [30, 45]. Hospital departments that can continually sense patients' needs are more effective in clinical communication, decision-support [7, 53], and the patient treatment process, thus responding effectively to patients' needs and wishes [15, 42]. A strong patient responding capability is likely to provide service flexibility, high-quality care, achieve patient satisfaction, and improve the accessibility of medical services [35, 43]. Following the literature on the widely adopted process-oriented perspective, it can be argued that patients' sensing capability effect on patient service performance is intermediated by patient responding capability [46]. Hence, this study defines the following:

**Hypothesis 4:** *Hospital departments' responding capability mediates the effect of sensing capability on the department's patient* service performance.

# 4 Research methods

## 4.1 Data

This survey was pretested on multiple occasions by five Master students and six medical practitioners and scholars to improve both the content and face validity of the survey items. The medical practitioners all had sufficient knowledge and experience to assess the survey items effectively to provide valuable improvement suggestions. The survey was anonymously administered to key informants within hospital departments. We assured the respondents that their entries would be treated confidentially, and we would only report outcomes on an aggregate level [47]. Our target population includes medical heads/chairs of the department, practicing doctors, and department managers.

Data were conveniently sampled from Dutch hospitals through 5 Master students' professional networks within Dutch hospitals. The data we collected between November 10$^{th,}$ 2019, to January 5$^{th,}$ 2020. This study uses 90 complete survey responses for final analyses. Within the obtained sample, 28.9% of the respondents work for a University medical center, 41,1% work for a specialized top clinical (training) hospital, and 30% work for general hospitals. Most survey respondents are medical heads/chairs of the department (51.1%)[1], 24.4% is a practicing doctor, 11.1% is department manager, while the remaining 13.3% of the respondents hold other positions such as specialized oncology nurses. ..

[1] 5 respondents claimed that they were team leads.

Finally, Harman's single-factor analysis was applied using exploratory factor analysis (in using IBM SPSS Statistics™ v24) to restrain, ex-post, possible common method bias [47]. Outcomes show that the current sample is not affected by method biases; no single factor is attributed to most of the variance.

### 4.2 Constructs and items

The selection of indicators was made based on previous empirical and validated work to increase the questions' internal validity and reliability. This study devised three core items from Khin and Ho [22] to measure *digital dynamic capability* and conceptualized *patient agility* as a higher-order dynamic capability comprising the dimensions '*patient sensing capability'* and '*patient responding capability'* [30, 32]. This study adopts five measures for each of these two capabilities, following Roberts and Grover [30]. This study builds upon the concept of IT-business value creation [24, 46] to conceptualize *patient service performance* (PSP). Thus, consistent with balanced evaluation perspectives, patient service performance is represented by three measures, i.e., enhanced quality, improved accessibility of medical services, and achieving patient satisfaction [24, 48, 49]. The constructs' items in the research model used a seven-point Likert scale ranging from 1: strongly disagree to 7: strongly agree. Following prior IS and management studies, we controlled for 'size' (full-time employees), operationalizing this measure using the natural log (i.e., log-normally distributed) and 'age' of the department (5-point Likert scale 1: 0–5years; 5: 25+ years). All measures are included in the Appendix, including the respective item-to-construct loadings ($\lambda$), mean values ($\mu$), and the standard deviations (*Std*.).

### 4.3 Analyses using Partial Least Squares

This study applied SmartPLS version 3.2.7. [50]. SmartPLS is a Structural Equation Modeling (SEM) tool that uses Partial Least Squares (PLS). PLS is used to estimate the research model and run its associated parameter estimates. A key reason for PLS's usage is that this current work focuses on predicting and articulating particular hypothesized relationships. Also, PLS allows flexibility concerning the assumptions on multivariate normality, the ability to run parameter estimates for smaller samples, and reduces the overall error associated with the model [51]. The sample size exceeds the minimum threshold to obtain stable PLS outcomes. This study uses a two-step approach to investigate PLS outcomes. First, the measurement model is assessed. Then, the hypotheses are tested using the outcomes of the structural model assessment.

## 5 Empirical results

### 5.1 Measurement model analysis

The internal consistency reliability, convergent validity, and measurement validity of the first-order latent constructs were first assessed. PLS outcomes show that all the construct-to-item loadings far exceeded a threshold of 0.70. Both the Cronbach's alpha (CA) and Composite reliability (CR) values are all above the threshold of 0.7, showcasing the latent constructs' reliability. Next, to assess convergent validity, the average variance extracted (AVE) was analyzed. This value captures the degree of variance explained by the latent construct while relating it to the amount of variance due to measurement error [50]. All AVE-values exceeded the threshold of 0.50 (the minimum was 0.66 for patient service performance). Three complimentary assessments were used to evaluate discriminant validity [52]. First, the Fornell-Larcker criterion, i.e., the AVE's square root, is compared with cross-correlation values. Outcomes show that each value is larger than cross-correlations [52]. Next, the cross-loadings between constructs are investigated. No single cross-loading exceeds a (correlation) difference of 0.20. Finally. This study investigated the heterotrait-monotrait ratio of correlations (HTMT) values [53]. HTMT values are all well below the upper bound of 0.90, once more confirming discriminant validity. The above outcomes suggest that the measurement model is both reliable and valid.

### 5.2 Structural model assessment and hypothesis testing

This study evaluates the Standardized Root Mean Square Residual (SRMR) value[2]to assess the model fit. The SRMR value calculates the difference between observed correlations and the model's implied correlations matrix [52]. The obtained SRMR of 0.069 is well below the conservative threshold of 0.08. The model's hypothesized relationships can now be estimated. This study investigated each hypothesized path's significance and the coefficient of determination ($R^2$), measuring the model's predictive power [52] to test the hypotheses. Also, the model's predictive power is assessed [52]. This study uses a non-parametric bootstrapping procedure (using 5000 replications) to obtain stable results and interpret their significance of the path coefficients between this study's key construct. Figure 1 shows the results of the structural model assessments.

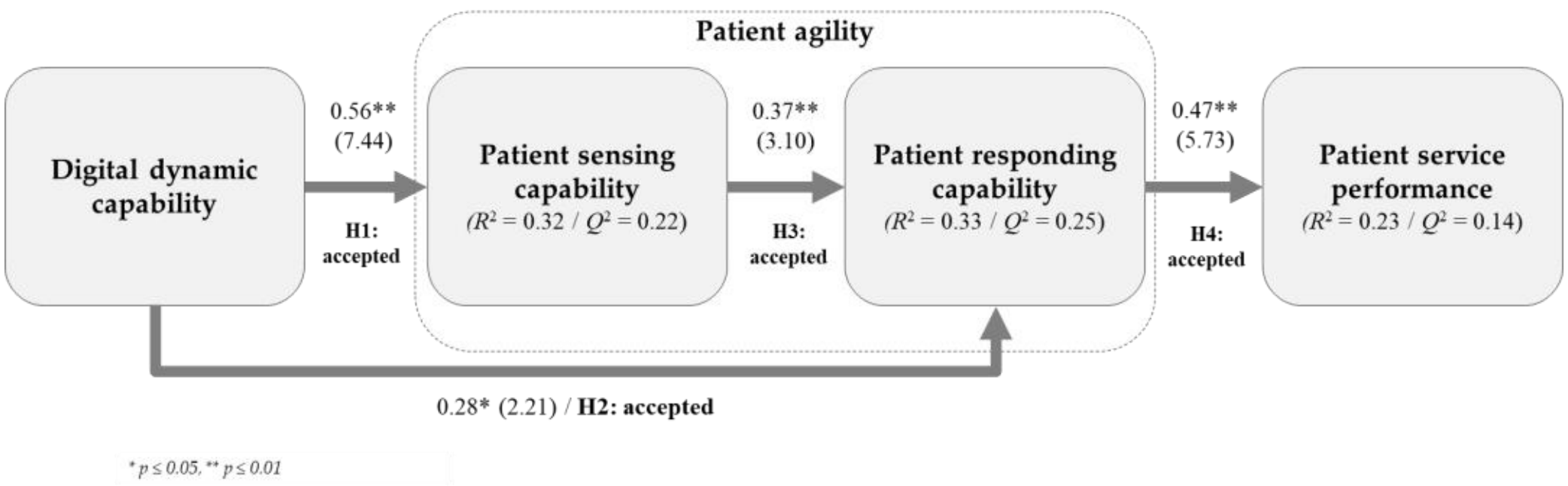

**Fig. 1.** Structural model assessment results using SmartPLS ($N$ = 90)

[2] This particular metric for model fit should still be interpreted with caution as it is not fully established in the PLS-SEM literature.

As can be seen from Figure 1, support was found for all the hypotheses. Digital dynamic capability positively influences patient sensing capability ($\beta = 0.56$; $t = 7.44$; $p < 0.0001$) and responding capability ($\beta = 0.38$; $t = 2.21$; $p = 0.03$). Patient sensing capability positively influences patient responding capability ($\beta = 0.37$; $t = 3.10$; $p = 0.002$). Finally, patient responding capability positively impacts patient service performance ($\beta = 0.47$; $t = 5.73$; $p < 0.0001$). The explained variance for sensing capability is 32% ($R^2 = .32$), for responding capability 33% ($R^2 = .33$), and for patient service performance ($R^2 = .23$). These amounts can be considered moderate effects. Specific mediation guidelines (Baron and Kenny 1986; Hair Jr et al. 2016; Hayes 2013) were followed to investigate the model's imposed mediation effects. Outcomes show that patient responding capability 'fully' mediates patient sensing capability on patient service performance [52, 54]. Concerning the first part of the model, patient sensing capability partially mediates the effect of digital dynamic capability. The included control variables shows non-significant effects: 'size' ($\beta = -0.01$, $t = 0.01$ $p = 0.92$), 'age' ($\beta = -0.01$, $t = 0.14$, $p = 0.89$). A subsequent blindfolding assessment for the endogenous latent constructs using Stone-Geisser values ($Q^2$) shows that the model has predictive power [52]. The $Q^2$ values far exceed 0, as can be seen in Figure 1.

## 6 Discussion and concluding remarks

This study aimed to understand better the crucial role of digital dynamic capability as an enabler of the hospital's ability to adequately 'sense' and 'respond' to patient needs and wishes and unfold how hospital departments can grasp the value of data-driven innovations. Modern hospitals need to ensure that their processes can meet the needs of an increasingly complex environment, especially now during the COVID-19 crisis. It is well known that it is essential to maintain strategic flexibility under acute conditions so that adequate digital options and sensing and responding behavior are exercised [45, 55]. However, many hospitals currently struggle in their digital transformation efforts in practice, and this process is usually painfully slow, with many hurdles to overcome.

This study makes two theoretical contributions. This study is the first to conceptualize patient agility and empirically validate that patient agility will enhance hospital departments' service performance. This study found support for this claim. It shows that digital dynamic capability enhances patient agility's conceptualized construct by sequentially enhancing patient sensing capability and patient responding capability. These results are coherent with previous work prompting that those hospital departments that invest and enhance their skills, competences, and knowledge in managing innovative digital technologies are better equipped to be responsive, innovative, and satisfy patients' needs [22, 24]. Also, this study adds to the growing body of knowledge on the degree to which digital capabilities and competencies contribute to organizational capabilities and benefits [33, 44]. The obtained insights are valuable as future research can consider these particular insights when investigating hospitals' IT-business value.

The outcomes of this study provide managerial implications in several ways. First, the outcomes are relevant for the healthcare sector now, as hospitals worldwide need to transform healthcare delivery processes using digital technologies and increase clinical productivity during the COVID-19 crisis [56]. Hospital departments need to develop the dynamic capabilities and direct IT investments to bring about the highest IT business value. Hence, departments should prioritize clinical initiatives by focusing on critical workflows and clinical process improvement opportunities, emphasizing patient agility. The department's digital dynamic capability is crucial in the development of new digital patient service solutions. So, hospital departments need to actively invest in the skills and competencies to manage new digital technologies like big data and predictive analytics and AI. Hospitals typically will need to overcome adoption (e.g., physician resistance), process and technology challenges to develop the department's ability to master digital technologies, drive digital transformations, and adopt new innovative services and products. Therefore, hospital decision-makers must deliberately pay attention to stakeholder involvement and provide appropriate tangible and intangible resources [57, 58].

The study's limitations are now addressed so that the discussion is put into a proper academic context. First, data were collected at a single point in time (cross-sectionally) and thus providing only a snapshot of the firm's well-being. Also, capability building and achieving patient service performance are typically part of a hospital department's long-term goals and strategic direction. Therefore, a longitudinal approach could be valuable in providing a richer understanding of the dynamics among this study's constructs. *Second*, all the collected data in this study is from the Dutch hospitals. Therefore, the study outcomes could be subject to specific Country, cultural, and (local) economic influences. Notwithstanding, a substantial amount of current scholarships comes from Western (North America and Western-Europe) countries. This study, therefore, fits into a broader class of studies.

## References

1. Curtright, J.W., S.C. Stolp-Smith, and E.S. Edell, Strategic performance management: development of a performance measurement system at the Mayo Clinic. Journal of Healthcare Management, 2000. 45(1): p. 58–68.
2. Ahovuo, J., et al., Process oriented organisation in the regional PACS environment. EuroPACS-MIR 2004 in the Enlarged Europe, 2004: p. 481–484
3. McGlynn, E., et al., The quality of health care delivered to adults in the United States. New England Journal of Medicine, 2003. 348(26): p. 2635–2645.
4. Chiasson, M., et al., Expanding multi-disciplinary approaches to healthcare information technologies: what does information systems offer medical informatics? International Journal of Medical Informatics, 2007. 76: p. S89–S97.
5. Lee, J., J.S. McCullough, and R.J. Town, The impact of health information technology on hospital productivity. The RAND Journal of Economics, 2013. 44(3): p. 545-568.
6. Van de Wetering, R., J. Versendaal, and P. Walraven. Examining the relationship between a hospital's IT infrastructure capability and digital capabilities: a resource-based perspective. In: Proceedings of the Twenty-fourth Americas Conference on Information Systems (AMCIS), AIS, New Orleans (2018).

7. Van de Wetering, R., IT-Enabled Clinical Decision Support: An Empirical Study on Antecedents and Mechanisms. Journal of Healthcare Engineering, 2018. 2018: p. 10.
8. Hendrikx, H., et al., Expectations and attitudes in eHealth: A survey among patients of Dutch private healthcare organizations. International Journal of Healthcare Management, 2013. 6(4): p. 263-268.
9. Kohli, R. and S.S.-L. Tan, Electronic health records: how can IS researchers contribute to transforming healthcare? Mis Quarterly, 2016. 40(3): p. 553-573.
10. Garg, A.X., et al., Effects of computerized clinical decision support systems on practitioner performance and patient outcomes: a systematic review. Jama, 2005. 293(10): p. 1223-1238.
11. Romano, M.J. and R.S. Stafford, Electronic health records and clinical decision support systems: impact on national ambulatory care quality. Archives of internal medicine, 2011. 171(10): p. 897-903.
12. Van de Wetering, R. Enhancing clinical decision support through information processing capabilities and strategic IT alignment. In: Proceedings of the 21st International Conference on Business Information Systems, Springer, Berlin, (2018).
13. Wang, Y., L. Kung, and T.A. Byrd, Big data analytics: Understanding its capabilities and potential benefits for healthcare organizations. Technological Forecasting and Social Change, 2018. 126: p. 3-13.
14. Prgomet, M., A. Georgiou, and J.I. Westbrook, The impact of mobile handheld technology on hospital physicians' work practices and patient care: a systematic review. Journal of the American Medical Informatics Association, 2009. 16(6): p. 792-801.
15. Li, W., et al., Integrated clinical pathway management for medical quality improvement–based on a semiotically inspired systems architecture. European Journal of Information Systems, 2014. 23(4): p. 400-417.
16. Andargoli, A.E., et al., Health information systems evaluation frameworks: a systematic review. International journal of medical informatics, 2017. 97: p. 195-209.
17. Van de Wetering, R. IT ambidexterity and patient agility: the mediating role of digital dynamic capability. In: Proceedings of the Twenty-Ninth European Conference on Information Systems (ECIS), AIS, Virtual Conference, (2021).
18. Mikalef, P., R. van de Wetering, and J. Krogstie, Building dynamic capabilities by leveraging big data analytics: The role of organizational inertia. Information & Management, 2020: p. 103412.
19. Teece, D., M. Peteraf, and S. Leih, Dynamic capabilities and organizational agility: Risk, uncertainty, and strategy in the innovation economy. California Management Review, 2016. 58(4): p. 13-35.
20. Van de Wetering, R., Dynamic Enterprise Architecture Capabilities and Organizational Benefits: An empirical mediation study. In: Proceedings of the Twenty-Eigth European Conference on Information Systems, AIS, Virtual Conference, (2020).
21. Van de Wetering, R., et al., The Impact of EA-Driven Dynamic Capabilities, Innovativeness, and Structure on Organizational Benefits: A Variance and fsQCA Perspective. Sustainability, 2021. 13(10): p. 5414.
22. Khin, S. and T.C. Ho, Digital technology, digital capability and organizational performance: A mediating role of digital innovation. International Journal of Innovation Science, 2019. 11(2): p. 177-195.

23. Teece, D.J., G. Pisano, and A. Shuen, Dynamic capabilities and strategic management. Strategic management journal, 1997. 18(7): p. 509-533.
24. Wu, L. and Y.-P. Hu, Examining knowledge management enabled performance for hospital professionals: A dynamic capability view and the mediating role of process capability. Journal of the Association for Information Systems, 2012. 13(12): p. 976.
25. Eisenhardt, K.M. and J.A. Martin, Dynamic capabilities: what are they? Strategic management journal, 2000. 21(10-11): p. 1105-1121.
26. Teece, D.J., Explicating dynamic capabilities: the nature and microfoundations of (sustainable) enterprise performance. Strategic management journal, 2007. 28(13): p. 1319-1350.
27. Wang, C.L. and P.K. Ahmed, Dynamic capabilities: A review and research agenda. International journal of management reviews, 2007. 9(1): p. 31-51.
28. Wilden, R. and S.P. Gudergan, The impact of dynamic capabilities on operational marketing and technological capabilities: investigating the role of environmental turbulence. Journal of the Academy of Marketing Science, 2015. 43(2): p. 181-199.
29. Tallon, P.P., et al., Information technology and the search for organizational agility: A systematic review with future research possibilities. The Journal of Strategic Information Systems, 2019. 28(2): p. 218-237.
30. Roberts, N. and V. Grover, Leveraging information technology infrastructure to facilitate a firm's customer agility and competitive activity: An empirical investigation. Journal of Management Information Systems, 2012. 28(4): p. 231-270.
31. Park, Y., O.A. El Sawy, and P.C. Fiss, The Role of Business Intelligence and Communication Technologies in Organizational Agility: A Configurational Approach. Journal of the Association for Information Systems, 2017. 18(9): p. 648-686.
32. Sambamurthy, V., A. Bharadwaj, and V. Grover, Shaping agility through digital options: Reconceptualizing the role of information technology in contemporary firms. MIS quarterly, 2003. 27(2): p. 237-263.
33. Chakravarty, A., R. Grewal, and V. Sambamurthy, Information technology competencies, organizational agility, and firm performance: Enabling and facilitating roles. Information systems research, 2013. 24(4): p. 976-997.
34. Lu, Y. and K. Ramamurthy, Understanding the link between information technology capability and organizational agility: An empirical examination. 2011. 35(4): p. 931-954.
35. Roberts, N. and V. Grover, Investigating firm's customer agility and firm performance: The importance of aligning sense and respond capabilities. Journal of Business Research, 2012. 65(5): p. 579-585.
36. Acur, N., et al., Exploring the impact of technological competence development on speed and NPD program performance. Journal of product innovation management, 2010. 27(6): p. 915-929.
37. Zhou, K.Z. and F. Wu, Technological capability, strategic flexibility, and product innovation. Strategic Management Journal, 2010. 31(5): p. 547-561.
38. Li, T.C. and Y.E. Chan, Dynamic information technology capability: Concept definition and framework development. The Journal of Strategic Information Systems, 2019. 28(4): p. 101575.

39. Božič, K. and V. Dimovski, Business intelligence and analytics use, innovation ambidexterity, and firm performance: A dynamic capabilities perspective. The Journal of Strategic Information Systems, 2019. 28(4): p. 101578.
40. Wang, Y., et al., Leveraging big data analytics to improve quality of care in healthcare organizations: A configurational perspective. British Journal of Management, 2019. 30(2): p. 362-388.
41. Coombs, J.E. and P.E. Bierly III, Measuring technological capability and performance. R&D Management, 2006. 36(4): p. 421-438.
42. Salge, T.O. and A. Vera, Hospital innovativeness and organizational performance: Evidence from English public acute care. Health Care Management Review, 2009. 34(1): p. 54-67.
43. Bradley, R., et al. An examination of the relationships among IT capability intentions, IT infrastructure integration and quality of care: A study in US hospitals. in 2012 45th Hawaii International Conference on System Sciences. 2012. IEEE.
44. Chen, Y., et al., IT capability and organizational performance: the roles of business process agility and environmental factors. European Journal of Information Systems, 2014. 23(3): p. 326-342.
45. Overby, E., A. Bharadwaj, and V. Sambamurthy, Enterprise agility and the enabling role of information technology. European Journal of Information Systems, 2006. 15(2): p. 120-131.
46. Schryen, G., Revisiting IS business value research: what we already know, what we still need to know, and how we can get there. European Journal of Information Systems, 2013. 22(2): p. 139-169.
47. Podsakoff, P.M., et al., Common method biases in behavioral research: A critical review of the literature and recommended remedies. Journal of applied psychology, 2003. 88(5): p. 879.
48. Chen, J.-S. and H.-T. Tsou, Performance effects of IT capability, service process innovation, and the mediating role of customer service. Journal of Engineering and Technology Management, 2012. 29(1): p. 71-94.
49. Setia, P., V. Venkatesh, and S. Joglekar, Leveraging digital technologies: How information quality leads to localized capabilities and customer service performance. Mis Quarterly, 2013. 37(2).
50. Ringle, C.M., S. Wende, and J.-M. Becker, SmartPLS 3. Boenningstedt: SmartPLS GmbH, http://www. smartpls. com, 2015.
51. Hair Jr, J.F., et al., Advanced issues in partial least squares structural equation modeling. 2017: SAGE Publications.
52. Hair Jr, J.F., et al., A primer on partial least squares structural equation modeling (PLS-SEM). 2016: Sage Publications.
53. Henseler, J., C.M. Ringle, and M. Sarstedt, A new criterion for assessing discriminant validity in variance-based structural equation modeling. Journal of the Academy of Marketing Science, 2015. 43(1): p. 115-135.
54. Hayes, A.F., Introduction to mediation, moderation, and conditional process analysis: A regression-based approach. 2013: Guilford Press.
55. D'Aveni, R.A., G.B. Dagnino, and K.G. Smith, The age of temporary advantage. Strategic management journal, 2010. 31(13): p. 1371-1385.

56. Keesara, S., A. Jonas, and K. Schulman, Covid-19 and health care's digital revolution. New England Journal of Medicine, 2020. 382(23): p. e82.
57. Gray, C.S., Seeking Meaningful Innovation: Lessons Learned Developing, Evaluating, and Implementing the Electronic Patient-Reported Outcome Tool. Journal of Medical Internet Research, 2020. 22(7): p. e17987.
58. Papoutsi, C., et al., Putting the social back into sociotechnical: Case studies of co-design in digital health. Journal of the American Medical Informatics Association, 2020.

## Appendix A: Constructs and items

| **Construct** | | **Measurement item** | *λ* | *μ* | ***Std.*** | ***Reliability statistics*** |
|---|---|---|---|---|---|---|
| *Digital dynamic capability* | *Please indicate the ability of your department to: (1. Strongly disagree–7. Strongly agree* | | | | | |
| | DDC1 | Responding to digital transformation | 0.886 | 4.33 | 1.56 | CA: 0.86<br>CR:0.91<br>AVE:0.78 |
| | DDC2 | Mastering the state-of-the-art digital technologies | 0.895 | 3.69 | 1.48 | |
| | DDC3 | Developing innovative patient services using digital technology | 0.856 | 4.74 | 1.63 | |
| *Sensing* | *Indicate the degree to which you agree or disagree with the following statements about whether the department can (1 – strongly disagree 7 – strongly agree)* | | | | | |
| | S1 | We continuously discover additional needs of our patients of which they are unaware | 0.89 | 4.10 | 1.66 | CA:0.89<br>CR:0.92<br>AVE:0.71 |
| | S2 | We extrapolate key trends for insights on what patients will need in the future | 0.77 | 4.43 | 1.63 | |
| | S3 | We continuously anticipate our patients' needs even before they are aware of them | 0.89 | 4.03 | 1.68 | |
| | S4 | We attempt to develop new ways of looking at patients and their needs | 0.79 | 4.72 | 1.52 | |
| | S5 | We sense our patient's needs even before they are aware of them | 0.86 | 3.94 | 1.66 | |
| *Responding* | R1 | We respond rapidly if something important happens with regard to our patients | 0.93 | 4.52 | 1.50 | CA:0.91<br>CR:0.93<br>AVE:0.89 |
| | R2 | We quickly implement our planned activities with regard to patients | 0.91 | 4.52 | 1.42 | |
| | R3 | We quickly react to fundamental changes with regard to our patients | 0.92 | 4.54 | 1.53 | |
| | R4 | When we identify a new patient need, we are quick to respond to it | 0.87 | 4.11 | 1.62 | |
| | R5 | We are fast to respond to changes in our patient's health service needs | 0.87 | 4.76 | 1.71 | |
| *PSP* | *We perform much better during the last 2 or 3 years than comparable departments from other hospitals in: (1. Strongly disagree–7. Strongly agree).* | | | | | |
| | PSP1 | Achieving patient satisfaction | 0.83 | 4.98 | 1.32 | CA:0.75<br>CR:0.85<br>AVE:0.66 |
| | PSP2 | Providing high-quality service | 0.85 | 5.28 | 1.25 | |
| | PSP3 | Improving the accessibility of medical services | 0.75 | 4.80 | 1.33 | |